\documentclass[reprint]{JASA}

\usepackage[table]{xcolor}
\colorlet{shadecolor}{gray!15}

\usepackage{graphicx}
\usepackage{amsmath,amssymb,graphicx}
\usepackage[]{algorithm2e}
\usepackage{tabularx}
\usepackage{wasysym}
\usepackage{color,soul}
\usepackage{balance}
\usepackage{booktabs}
\usepackage{multirow}
\usepackage{siunitx}
\usepackage{hyperref}
\def\H{{\mathsf H}}

\def\CC{{\mathbb C}}
\def\RR{{\mathbb R}}
\def\ZZ{{\mathbb Z}}
\usepackage{makecell}
\usepackage{textgreek}
\usepackage{enumitem}
\usepackage{flushend}
\usepackage{makecell}

\usepackage{amssymb} %
\usepackage{pifont} %
\newcommand{\cmark}{\ding{51}}%
\newcommand{\xmark}{\ding{55}}%

\usepackage{arydshln}
\newcommand{\ZQHL}[1]{#1} %

\begin{document}

\title[]{ctPuLSE: Close-Talk, and Pseudo-Label Based Far-Field, Speech Enhancement}
\author{Zhong-Qiu Wang}
\affiliation{Department of Computer Science and Engineering, Southern University of Science and Technology, Shenzhen 518055, Guangdong, P.R. China}
\email{wang.zhongqiu41@gmail.com or wangzq3@sustech.edu.cn}

\preprint{Wang, JASA}	%

\date{\today} 

\begin{abstract}
The current dominant approach for neural speech enhancement is via purely-supervised deep learning on simulated pairs of far-field noisy-reverberant speech (i.e., mixtures) and clean speech.
The trained models, however, often exhibit limited generalizability to real-recorded mixtures.
To deal with this, this paper investigates training enhancement models directly on real mixtures.
However, a major difficulty challenging this approach is that, since the clean speech of real mixtures is unavailable, there lacks a good supervision for real mixtures.
In this context, assuming that a training set consisting of real-recorded pairs of close-talk and far-field mixtures is available, we propose to address this difficulty via close-talk speech enhancement, where an enhancement model is first trained on simulated mixtures to enhance real-recorded close-talk mixtures and the estimated close-talk speech can then be utilized as a supervision (i.e., pseudo-label) for training far-field speech enhancement models directly on the paired real-recorded far-field mixtures.
We name the proposed system \textit{ctPuLSE}.
Evaluation results on the popular CHiME-4 dataset show that ctPuLSE can derive high-quality pseudo-labels and yield far-field speech enhancement models with strong generalizability to real data.
\end{abstract}

\maketitle

\section{Introduction}

Dramatic progress has been made in speech enhancement \cite{WDLreview}, thanks to the rapid development of deep learning.
The current dominant approach is based on supervised deep learning, where paired noisy-reverberant speech (i.e., mixtures) and clean speech are simulated and utilized to train deep neural networks (DNN) to predict the clean speech based on its paired mixtures in a purely-supervised, discriminative way \cite{WDLreview, Zheng2023sixty, Araki2025}.
The trained models, however, often exhibit limited generalizability to real-recorded mixtures \cite{Wang2024SuperM2M, Pandey2020CrossCorpus, Zhang2021ClosingGap, Tzinis2022REMIXT, Tzinis2022AudioScopeV2, Cox2023ClaritySP2023, Leglaive2023CHiME7UDASE, cornell23_chime7, HaebUmbach2019SPM, Zhang2023USES, Zhang2024USES2}, mainly because the simulated training data is typically, and in many cases inevitably, mismatched with real test data.

To improve the generalizability, this paper investigates training models directly on real-recorded, target-domain mixtures.
This approach, however, cannot be straightforwardly realized, since the underlying clean speech is not available for real mixtures, unlike simulated mixtures, where the paired clean speech can be readily available through room simulation and can serve as a fine-grained supervision at the sample level for model training.
The key to enable successful training on real mixtures, we think, is to figure out a mechanism that can reliably compute a high-quality supervision signal (i.e., pseudo-label or pseudo-target) for real-recorded mixtures.

\begin{figure}[t]
  \centering 
  \includegraphics[width=8.5cm]{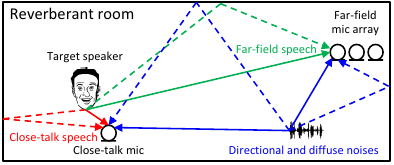}
  \caption{
  Scenario illustration.
  Close-talk mixture consists of close-talk speech and noises, and far-field mixture consists of far-field speech and noises.
  Best viewed in color.
  }
  \label{physical_model_figure}
\end{figure}

One possible way, we propose, is to leverage close-talk mixtures.
During data collection, besides using far-field microphones to record target speech, a close-talk microphone is often placed near the target speaker to record the close-talk speech at the same time.
See Fig. \ref{physical_model_figure} for an illustration.
The recorded close-talk mixture, while the target speaker is talking, usually has a much higher SNR of the target speaker than any far-field mixture, simply due to the very short distance from the target speaker to its close-talk microphone.
Although the close-talk mixture is typically not perfectly clean, as non-target signals (such as environmental noises, room reverberation and competing speech) could also be picked up by the close-talk microphone, it usually exhibits a very high input SNR and hence could be utilized to compute a reliable, high-quality supervision for real-recorded far-field mixtures.

With this understanding, this paper first investigates \textit{close-talk speech enhancement}, a task which aims at enhancing close-talk mixtures and estimating close-talk speech.
Solving this task could enable many applications.
One of them, which is investigated in this paper, is that, based on the estimated close-talk speech, a pseudo-label can be derived for each real-recorded far-field mixture and used as a supervision to train supervised models directly on real-recorded far-field mixtures, thereby potentially realizing better generalizability to real data.
We summarize the contributions of this paper as follows:
\begin{itemize}[leftmargin=*,noitemsep,topsep=0pt]
\item For the purpose of achieving better far-field speech enhancement, we propose to first investigate close-talk speech enhancement (CTSE), which aims at enhancing close-talk mixtures to estimate close-talk speech.
Although CTSE is a particular form of speech enhancement (i.e., in close-talk conditions) and has been studied for other purposes \cite{Jiang2013, Jiang2016,Tan2021CTSE}, we point out that it is a valuable task that could yield high-quality pseudo-labels for real-recorded data, and is hence worth investigating.
\item We propose ctPuLSE, a pseudo-label based approach for far-field speech enhancement, where estimated close-talk speech is utilized to derive a supervision for training supervised enhancement models directly based on real far-field mixtures.
\item Following SuperM2M \cite{Wang2024SuperM2M}, an earlier algorithm that trains enhancement models by alternating between supervised and un-/weakly-supervised learning, we propose a co-learning algorithm that trains the same enhancement model using both simulated and real data, where the pseudo-label of real data is derived based on the estimated close-talk speech.
This way, the model can also learn from massive amount of simulated training data, which can be easily simulated and can be very helpful when the real training data is scarce.
Compared with SuperM2M, we observe that ctPuLSE obtains comparable robust automatic speech recognition (ASR) performance, while much better enhancement performance.
\end{itemize}
Although ctPuLSE is simple and straightforward, it obtains strong robust ASR and speech enhancement performance on the public CHiME-4 dataset \cite{Barker2015,Vincent2016CHiME4Analysis,Barker2017CHiME3}, the most popular benchmark to date in robust ASR and speech enhancement.
The evaluation results suggest that ctPuLSE can effectively train enhancement models on real-recorded far-field mixtures, and can yield better generalizability to real data than purely-supervised models trained only on simulated data.
A sound demo is provided at \url{https://zqwang7.github.io/demos/ctPuLSE_demo/index.html}.

\section{Related Work}\label{related_work}

ctPuLSE is related to earlier works in four key aspects.

\subsection{Generalizability of Supervised Enhancement Models}

Improving the generalizability of supervised learning based speech enhancement models (trained on simulated data) to real-recorded data has received decade-long research interests.
The current dominant approach \cite{Chen2017b, Chen2016c, WDLreview, Zhang2023USES, Zhang2024USES2} is to simulate massive amount of training data to cover as many variations (that could happen in real-recorded test data) as possible.
However, the generalizability is often limited by the current simulation techniques, which usually cannot simulate mixtures as realistic as real-recorded mixtures.
This can be observed from recent speech enhancement/separation and robust ASR challenges such as CHiME-\{4,5,6,7\} \cite{Vincent2016CHiME4Analysis,Watanabe2020CHiME6,cornell23_chime7}, AliMeeting \cite{Yu2022M2MeT}, MISP \cite{Wu2024MISP2023}, and Clarity \cite{Cox2023ClaritySP2023}, where using the immediate outputs from DNN-based enhancement or separation models trained on simulated mixtures for robust ASR and human hearing has had limited successes \cite{HaebUmbach2019SPM, Haeb-Umbach2020}.
Different from this stream of research, this paper investigates training enhancement models directly on real-recorded data to improve the generalizability.

\subsection{Leveraging Close-Talk Mixtures for Speech Enhancement}

Leveraging real-recorded close-talk mixtures as a weak supervision to train speech enhancement and speaker separation models directly on real-recorded far-field mixtures has attracted research interests recently.
A representative algorithm in this direction is SuperM2M \cite{Wang2024SuperM2M}, which builds upon a weakly-supervised speaker separation algorithm named mixture to mixture (M2M) \cite{Wang2024M2MSPL} and an unsupervised speaker separation algorithm named UNSSOR \cite{Wang2023UNSSOR}.
M2M trains enhancement models to produce a speech estimate and a noise estimate such that the two estimates can be linearly filtered to recover observed mixtures, and SuperM2M, building upon M2M, leverages supervised learning on simulated mixtures to improve M2M.  
Although SuperM2M has shown strong potential for neural speech enhancement and robust ASR, we often observe that SuperM2M cannot sufficiently suppress non-target signals, likely because its loss function is defined on reconstructed mixtures (i.e., the summation of linearly-filtered source estimates), rather than directly on source estimates.
In comparison, the loss function in ctPuLSE, which we will show, is defined based on pseudo-labels derived from close-talk mixtures.
As long as the pseudo-labels are high-quality, we can reasonably expect that the suppression of non-target signals by ctPuLSE would be more sufficient and aggressive.

On the other hand, there are studies \cite{Wu2024MISP2023} using oracle speaker-activity timestamps and pre-trained models (e.g., DNSMOS \cite{K.A.Reddy2021DNSMOS}) to select segments of close-talk mixtures that are almost clean, and then using the selected segments of mixtures to synthesize far-field mixtures for training supervised learning based models.
However, the training data is still simulated, and the models are not trained on real data.

\subsection{Pseudo-Label Based Speech Enhancement}

There have been studies adapting pre-trained speech enhancement models to target domains via pseudo-labeling.
RemixIT \cite{Tzinis2022REMIXT}, a representative algorithm in this direction, first uses a pre-trained enhancement model (named \textit{teacher}) to enhance target-domain real-recorded mixtures (and generate pseudo-labels), and then another enhancement model (named \textit{student}) is trained in a supervised way to estimate the generated pseudo-labels.
The teacher and student models are designed to update continuously and iteratively to gradually improve each other.
Another recent study in this direction is SSST \cite{Frenkel2024}, which, building upon RemixIT, introduces an adversarial training algorithm to learn domain-invariant hidden representations and proposes a data-selection mechanism to pick a subset of target-domain mixtures whose pseudo-labels are sufficiently reliable for RemixIT-style training.

Similarly to those studies, ctPuLSE leverages pseudo-labeling as well, but it has a very different problem setup, where paired close-talk and far-field mixtures are assumed available for model training.
In this setup, much better pseudo-labels could potentially be computed for far-field mixtures, simply due to the innate high input SNR of close-talk mixtures.
This potential could make ctPuLSE a more attractive solution for practical product development, as long as the paired close-talk mixtures can be recorded while collecting far-field mixtures for training.

\ZQHL{
\subsection{Relations to DAPS Dataset}

A dataset named DAPS \cite{J.Mysore2015} is proposed for the task of transforming speech recorded on common consumer (and possibly band-limited) devices in real-world noisy-reverberant environments into professional production quality speech.
A loud speaker is used to play back clean speech in real-world environments, and the speech is then recorded by consumer devices to create pairs of clean and degraded speech signals for model training.
Although the DAPS paper proposes a task similar to the one in this study, it proposes a new dataset only rather than an actual algorithm.
On the other hand, in our study the close-talk mixture is recorded by attaching a close-talk microphone to the target talker and hence would inevitably capture some interference signals, and the pseudo-label is then derived based on the close-talk mixture.
Differently, DAPS uses played-back clean speech, which is used to derive pseudo-labels.
}

\subsection{Close-Talk Speech Enhancement}

Our study leverages close-talk speech enhancement to derive high-quality pseudo-labels for far-field speech enhancement.
There are existing studies \cite{Jiang2013, Jiang2016, Tan2021CTSE} on close-talk speech enhancement, but for different purposes.
For example, in \cite{Jiang2013, Jiang2016}, the task is to enhance the target speech captured by a close-talk microphone with the help of an ear-mounted microphone; and in \cite{Tan2021CTSE}, the task is to enhance close-talk speech when users talk to dual-microphone mobile phones at a very short distance.
Their purpose is not on training speech enhancement models on real data and improving the generalizability to real data.

\section{Physical Model and Approach Overview}

In a noisy-reverberant environment with a $P$-microphone far-field microphone array and a single target speaker wearing a close-talk microphone (see Fig. \ref{physical_model_figure} for an illustration), the physical model of the recorded close-talk mixture and each far-field mixture can be respectively formulated, in the short-time Fourier transform (STFT) domain, as follows:  
\begin{align}
Y_0(t,f) = X_0(t,f) + V_0(t,f), \label{phymodel_ct_raw} \\
Y_p(t,f) = X_p(t,f) + V_p(t,f), \label{phymodel_ff_raw}
\end{align}
where the subscript $p\in \{1,\dots,P\}$ indexes the $P$ far-field microphones, subscript $0$ indexes the close-talk microphone, $t\in\{0,\dots,T-1\}$ indexes $T$ frames, and $f\in\{0,\dots,F-1\}$ indexes $F$ frequency bins.
In (\ref{phymodel_ct_raw}), $Y_0(t,f)$, $X_0(t,f)$, and $V_0(t,f)$ respectively denote the STFT coefficients of the close-talk mixture, close-talk speech, and non-speech signals (such as environmental noises) captured by the close-talk microphone at time $t$ and frequency $f$.
Similarly, in (\ref{phymodel_ff_raw}), $Y_p(t,f)$, $X_p(t,f)$, and $V_p(t,f)$ are respectively the STFT coefficients of the far-field mixture, far-field speech, and non-speech signals captured by far-field microphone $p$ at time $t$ and frequency $f$.
In the rest of this paper, when dropping the indices $p$, $t$ and $f$, we refer to the corresponding spectrograms.
In this paper, $V$ is assumed to contain an unknown number of strong, non-stationary diffuse and directional noises sources.

\begin{figure}
  \centering  
  \includegraphics[width=8.75cm]{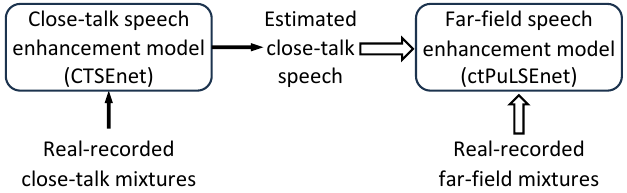}
  \caption{
    ctPuLSE illustration.
    A CTSEnet is first trained on simulated mixtures for close-talk speech enhancement.
    It is then used to enhance real-recorded close-talk mixtures and obtain estimated close-talk speech.
    Next, a ctPuLSEnet is trained for far-field speech enhancement, based on the estimated close-talk speech and its paired real-recorded far-field mixtures.
  }
  \label{ctPuLSE_figure}
\end{figure}

Fig. \ref{ctPuLSE_figure} illustrates the proposed ctPuLSE algorithm.
Given a set of simulated pairs of clean speech and mixtures as well as a set of real-recorded pairs of close-talk and far-field mixtures for training, the proposed system consists of three steps:
(a) we train a speech enhancement model (denoted as \textbf{CTSEnet}) based on the simulated training mixtures for close-talk speech enhancement;
(b) we apply the trained CTSEnet to enhance each real-recorded close-talk mixture and obtain an estimated close-talk speech;
and (c) we leverage the estimated close-talk speech as pseudo-labels to train a far-field speech enhancement model (denoted as \textbf{ctPuLSEnet}) on the paired real-recorded far-field mixtures.

Next, we detail CTSEnet and ctPuLSEnet.

\section{CTSEnet}\label{close_talk_SE}

Based on a set of simulated training mixtures, we train CTSEnet via single-microphone complex spectral mapping \cite{Tan2020, Wang2020aCSMCHiME4} to predict the real and imaginary (RI) components of target speech $X_a$ based on the RI components of the input mixture $Y_a$, where, depending on the application scenarios, the microphone index $a$ can index all or a subset of the simulated microphones in the training data.
Following \cite{Wang2021compensation}, we define the loss function on the RI components and magnitude of the DNN estimate $\hat{X}_a$:
\begin{align}\label{loss_CTSE}
\mathcal{L}_{X,a}^{\text{simu}} = \frac{ \sum\nolimits_{t,f} \mathcal{F}\Big(\hat{X}_a(t,f),X_a(t,f)\Big) }{\sum\nolimits_{t',f'} \big| X_a(t',f') \big|},
\end{align}
\begin{align}\label{loss_CTSE_F}
\mathcal{F}\Big(\hat{X}_a(t,f), X_a(t,f)\Big) = &\Big| \mathcal{R}\big(\hat{X}_a(t,f)\big) -  \mathcal{R}\big(X_a(t,f)\big) \Big|, \nonumber \\
&+\Big| \mathcal{I}\big(\hat{X}_a(t,f)\big) -  \mathcal{I}\big(X_a(t,f)\big) \Big|, \nonumber \\
&+\Big| |\hat{X}_a(t,f)| -  |X_a(t,f)| \Big|,
\end{align}
where $|\cdot|$ computes the magnitude or absolute value, $\mathcal{R}(\cdot)$ and $\mathcal{I}(\cdot)$ respectively extract the RI components, and the denominator in (\ref{loss_CTSE}) balances the loss values across different training mixtures.
Other configurations of CTSEnet are detailed later in Section \ref{training_setup} and \ref{configuration}.

\section{ctPuLSEnet}\label{PuLSEnet}

Once a CTSEnet has been trained, we apply it to enhance each monaural real-recorded close-talk mixture to obtain an estimate of the close-talk speech, denoted as $\hat{X}_0^{\text{CTSE}}$, which is then leveraged as a pseudo-label to train ctPuLSEnet on real-recorded far-field mixtures.
This section describes the DNN configurations, loss functions, and a co-learning algorithm that trains ctPuLSEnet on both simulated and real mixtures.

\subsection{DNN Configurations}\label{PuLSE_dnn_config}

ctPuLSEnet is trained via single- or multi-microphone complex spectral mapping \cite{Wang2020dMCCSMconference, Wang2020css, Tan2022NSF} to predict the RI components of target speech at a designated reference microphone $q\in\{1,\dots,P\}$ based on the RI components of stacked input mixtures.
In monaural cases, the DNN is trained to predict the target speech $X_q$ based on input mixture $Y_q$, and in multi-channel cases, it is trained to predict $X_q$ based on all the input mixtures stacked in a fixed microphone order.
We denote the DNN estimate as $\hat{X}_q$.
Besides predicting $X_q$, the DNN can also be trained to additionally predict non-speech signals $V_q$.
We observe that this form of multi-task learning can improve enhancement and robust ASR in our experiments.

Next, we describe the loss functions for the speech estimate $\hat{X}_q$ in Section \ref{PuLSE_loss} and for the noise estimate $\hat{V}_q$ in \ref{PuLSE_colearning}.
Other DNN setups are detailed in Section \ref{training_setup} and \ref{configuration}.

\subsection{Loss Functions Based on Estimated Close-Talk Speech}\label{PuLSE_loss}

Assuming that the close-talk microphone and far-field microphones are approximately time-synchronized, we directly leverage $\hat{X}_0^{\text{CTSE}}$ as the pseudo-label for each far-field mixture $Y_p$.
Since $\hat{X}_0^{\text{CTSE}}$ is time- and gain-aligned to close-talk speech $X_0$ rather than to far-field speech $X_p$ in far-field mixture $Y_p$, special care is needed to account for the time delay and gain differences between $\hat{X}_0^{\text{CTSE}}$ and $X_p$.

In this context, we propose to first linearly filter the DNN estimate $\hat{X}_p$ at each frequency to align it to $\hat{X}_0^{\text{CTSE}}$ before loss computation:
\begin{align}\label{loss_PuLSE}
\mathcal{L}_{X}^{\text{real}} = \frac{ \sum\nolimits_{t,f} \mathcal{F}\Big( \hat{X}_{p\rightarrow 0}(t,f), \hat{X}_0^{\text{CTSE}}(t,f) \Big) }{\sum\nolimits_{t',f'} \big| \hat{X}_0^{\text{CTSE}}(t',f') \big|},
\end{align}
\begin{align}\label{loss_lp}
\hat{X}_{p\rightarrow 0}(t,f) = \mathbf{\hat{g}}_p(f)^\H \widetilde{\hat{\mathbf{X}}}_p(t,f)
\end{align}
where $\widetilde{\hat{\mathbf{X}}}_p(t,f)=\big[\hat{X}_p(t-I+1,f),\dots,\hat{X}_p(t,f),\dots,\hat{X}_p(t+J,f)\big]\in\CC^{I+J}$ stacks a window of T-F units, $\mathbf{\hat{g}}_p(f)\in \CC^{I+J}$ denotes an estimated multi-tap linear filter to be described in (\ref{fcp_proj_mixture}), and $\mathcal{F}(\cdot,\cdot)$ is a distance metric defined in (\ref{loss_CTSE_F}).
Following the forward convolutive prediction (FCP) algorithm \cite{Wang2021FCPjournal}, we estimate $\mathbf{g}_p(f)$ by linearly projecting the DNN estimate $\hat{X}_p(\cdot,f)$ to pseudo-label $\hat{X}_0^{\text{CTSE}}(\cdot,f)$ at each frequency $f$:
\begin{align}\label{fcp_proj_mixture}
\hat{\mathbf{g}}_p(f)  =
\underset{\mathbf{g}_p(f)}{\text{argmin}}
\sum\limits_t \Big| \hat{X}_0^{\text{CTSE}}(t,f) - \mathbf{g}_p(f)^\H \widetilde{\hat{\mathbf{X}}}_p(t,f) \Big|^2.
\end{align}
This is a quadratic problem, which has a closed-form solution.
Next, we plug the closed-form solution into (\ref{loss_lp}), compute the loss in (\ref{loss_PuLSE}), and train the DNN.

An alternative way for linear filtering is to estimate the filter in the time domain, following the ideas behind CI-SDR \cite{Boeddeker2021CISDR}.
In detail, we compute a time-domain multi-tap Wiener filter to align $\hat{x}_p$ to $\hat{x}_0^{\text{CTSE}}$, where $\hat{x}_p=\text{iSTFT}(\hat{X}_p)$ is the re-synthesized time-domain signal of $\hat{X}_p$ obtained via inverse STFT (iSTFT), and similarly $\hat{x}_0^{\text{CTSE}}=\text{iSTFT}(\hat{X}_0^{\text{CTSE}})$.
The filter is computed by solving the following problem:
\begin{align}\label{WF_filtering}
\hat{h}_p = \underset{h_p}{\text{argmin}}\sum_n \Big| \hat{x}_0^{\text{CTSE}}[n] - \big(h_p * \hat{x}_p\big)[n] \Big|^2,
\end{align}
where $n$ indexes time-domain samples, $*$ denotes linear convolution, and $h_p\in\RR^{K+1+K}$ denotes a $(K+1+K)$-tap linear filter with $K$ past and $K$ future taps. 
(\ref{WF_filtering}) is also a quadratic linear regression problem, where a closed-form solution can be readily computed.
After that, we compute the STFT spectrogram of the filtered time-domain DNN estimate
\begin{align}\label{loss_lp_time}
\hat{X}_{p\rightarrow 0}=\text{STFT}\big(\hat{h}_p*\hat{x}_p\big),
\end{align}
which is then plugged into (\ref{loss_PuLSE}) to compute the loss for training.

\subsection{Addressing Time-Synchronization Issues in Pre-Processiing}\label{sync_main_body}

In the previous subseciton, when performing linear filtering for loss computation, we assume that far-field and close-talk microphones are reasonably time-synchronized.
In practical data collection, the close-talk microphone and far-field microphone array are usually managed and processed by two different devices.
Although the microphones on each device are typically synchronized, there could be synchronization errors between the microphones on different devices.
Later in Section \ref{sync_ct_ff}, we will slightly modify the classic GCC-PHAT algorithm \cite{DiBiase2001, Wang2018hLocalization} to roughly synchronize the close-talk microphone and far-field array.
This technique is utilized as a pre-processing operation prior to training.

\subsection{Co-Learning on Simulated and Real Mixtures}\label{PuLSE_colearning}

\begin{figure}[t]
  \centering  
  \includegraphics[width=8.5cm]{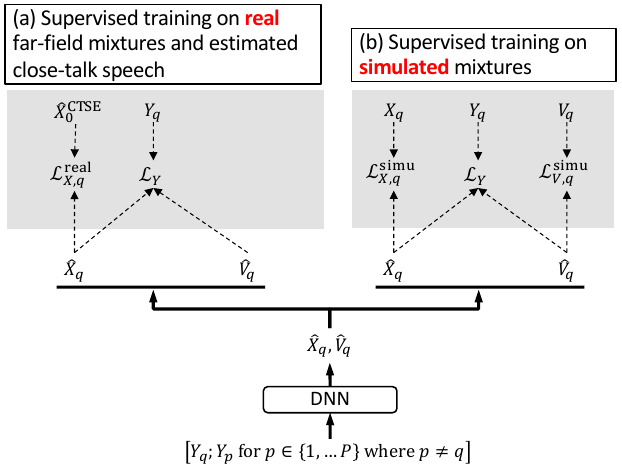}
  \caption{
    Illustration of co-learning, which trains the same model via supervised learning based on (a) real-recorded far-field mixtures and their pseudo-labels derived from CTSEnet; and (b) simulated mixtures.
  }
  \label{colearning_figure}
\end{figure}

In practical application scenarios, the amount of real data is typically scarce, as it is often effort-consuming to collect real-recorded close-talk and far-field mixture pairs.
In this case, even if ctPuLSEnet can be trained on real data, the performance is often limited, simply due to the limited amount of training data.
On the other hand, simulated data can be easily and massively generated through simulation.
In this case, following the SuperM2M algorithm \cite{Wang2024SuperM2M}, which combines supervised learning on simulated data with unsupervised or weakly-supervised learning on real data, we propose to train ctPuLSEnet on both simulated and real data.
See Fig. \ref{colearning_figure} for an illustration, where supervised learning on simulated data is shown in Fig. \ref{colearning_figure}(b).

Specifically, at each training step, we randomly sample either a mini-batch of real far-field mixtures and their pseudo-labels computed by CTSEnet or a mini-batch of simulated far-field mixtures (where the clean speech is available) to train the same DNN.
If the sampled mini-batch is real, the loss function can be the $\mathcal{L}_{X,q}^{\text{real}}$ loss defined in (\ref{loss_PuLSE}), while if the sampled mini-batch is simulated, the loss can be $\mathcal{L}_{X,q}^{\text{simu}}$ in (\ref{loss_CTSE}).

Alternatively, we can train ctPuLSEnet to not only predict target speech but also non-target signals via multi-task learning, as is described earlier in Section \ref{PuLSE_dnn_config}.
In this case, the loss on simulated data can be
\begin{align}\label{loss_colearning_simu}
\mathcal{L}_{q}^{\text{simu}} = \mathcal{L}_{X,q}^{\text{simu}} + \mathcal{L}_{V,q}^{\text{simu}} + \mathcal{L}_{Y,q},
\end{align}
where $\mathcal{L}_{V,q}^{\text{simu}}$ denotes the loss on estimated noise $\hat{V}_q$ and is defined, by following $\mathcal{L}_{X,q}^{\text{simu}}$ in (\ref{loss_CTSE}), as
\begin{align}\label{loss_colearning_simu_V}
\mathcal{L}_{V,q}^{\text{simu}} = \frac{ \sum\nolimits_{t,f} \mathcal{F}\Big(\hat{V}_q(t,f),V_q(t,f)\Big) }{\sum\nolimits_{t',f'} \big| V_q(t',f') \big|},
\end{align}
and $\mathcal{L}_{Y,q}$ is a mixture-constraint loss \cite{Wang2022GridNetjournal} encouraging the speech and noise estimates to sum up to the observed mixture:
\begin{align}\label{loss_colearning_simu_Y}
\mathcal{L}_{Y,q} = \frac{ \sum\nolimits_{t,f} \mathcal{F}\Big(\hat{X}_q(t,f)+\hat{V}_q(t,f),Y_q(t,f)\Big) }{\sum\nolimits_{t',f'} \big| Y_q(t',f') \big|}.
\end{align}
Accordingly, the loss on real data can be modified to
\begin{align}\label{loss_colearning_real}
\mathcal{L}_q^{\text{real}} = \mathcal{L}_{X,q}^{\text{real}} + \mathcal{L}_{Y,q}.
\end{align}
where $\mathcal{L}_{X,q}^{\text{real}}$ is defined in (\ref{loss_PuLSE}).

Notice that in each of the loss functions in (\ref{loss_CTSE}), (\ref{loss_PuLSE}), (\ref{loss_colearning_simu_V}) and (\ref{loss_colearning_simu_Y}), a normalization term is used in the denominator to balance the loss with the others.
This is particularly useful for $\mathcal{L}_{X}^{\text{real}}$ in (\ref{loss_PuLSE}), as the close-talk speech estimated based on the close-talk mixture (i.e., $\hat{X}_0^{\text{CTSE}}$) could have a gain level very different from far-field mixtures.

We can also add a weighting term $\alpha\in\RR_{>0}$ between the loss on simulated data and the loss on real data to balance their importance.
The overall loss is defined as
\begin{equation}
\mathcal{L}_q = \left\{
\begin{aligned}
&\alpha \times \mathcal{L}_q^{\text{simu}}, \text{for mini-batches of simulated data;} \\
&\mathcal{L}_q^{\text{real}}, \text{for mini-batches of real data.}
\end{aligned}
\right.
\label{final_loss_simu_real}
\end{equation}

\begin{table*}[]
\scriptsize
\centering
\sisetup{table-format=2.2,round-mode=places,round-precision=2,table-number-alignment = center,detect-weight=true,detect-inline-weight=math}
\caption{Number of utterances in CHiME-4}
\label{chime4_number_mixtures}
\setlength{\tabcolsep}{5pt}
\begin{tabular}{
c %
c %
c %
ccc
}
\toprule
Type & Close-talk or far-field? & \#mics & {Training Set} & {Validation Set} & {Test Set} \\
\midrule
SIMU & Far-field & 6 & $7,138$ ($\sim$$15.1$ h) & $1,640$ ($\sim$$2.9$ h) & $1,320$ ($\sim$$2.3$ h) \\
SIMU & Close-talk & {-} & {N/A} & {N/A} & {N/A} \\
\midrule
REAL & Far-field & 6 & $1,600$ ($\sim$$2.7$ h) & $1,640$ ($\sim$$2.7$ h) & $1,320$ ($\sim$$2.2$ h) \\
REAL & Close-talk & 1 & $1,600$ ($\sim$$2.7$ h) & $1,640$ ($\sim$$2.7$ h) & $1,320$ ($\sim$$2.2$ h) \\
\bottomrule
\end{tabular}
\end{table*}

\section{Experimental Setup}\label{setup}

Based on real-recorded mixtures, the main objective of our experiments is to demonstrate whether ctPuLSEnet can achieve better enhancement on real-recorded far-field mixtures than purely-supervised learning based approaches, which can only train enhancement models on simulated data.
Bearing this objective in mind, we validate the proposed algorithms on the CHiME-4 dataset \cite{Barker2015,Vincent2016CHiME4Analysis,Barker2017CHiME3}, the most popular benchmark to date in robust ASR and speech enhancement.
It contains both real-recorded and simulated mixtures for both training and testing.
Since, for real mixtures, the clean speech is not available for evaluation, we mainly check whether the enhanced speech can yield better ASR performance, considering that ASR scores can reflect speech intelligibility and indicate the degrees of speech distortion.
The ASR evaluation pipeline is shown in Fig. \ref{enh_asr_pipeline_figure}, where enhanced close-talk or far-field speech is directly fed to a strong pre-trained ASR system for recognition.

The rest of this section describes the CHiME-4 dataset, synchronization of close-talk and far-field mixtures, setup for ASR evaluation, miscellaneous system configurations, baselines for comparison, and evaluation metrics.

\subsection{CHiME-4 Dataset}\label{chime4_description}

CHiME-4 \cite{Barker2015, Vincent2016CHiME4Analysis, Barker2017CHiME3} is a major benchmark for evaluating far-field speech recognition and enhancement algorithms.
The far-field recording device is a tablet mounted with six microphones, with the second microphone placed on the rear and the other five facing front.
During data collection, the target speaker hand-holds the tablet and reads text prompts shown on the screen of the tablet.
The target speaker wears a close-talk microphone so that a monaural close-talk mixture can be recorded along with each six-channel far-field mixture.

The mixtures are recorded in four representative daily environments (including buses, cafeteria, pedestrian areas and streets), where multiple strong, non-stationary directional and diffuse noises can naturally co-exist.
In CHiME-4, the room reverberation is weak, and the major challenge is in how to deal with the multi-source non-stationary noise signals.

The number of mixtures in CHiME-4 is summarized in Table \ref{chime4_number_mixtures}.
Besides real mixtures, CHiME-4 also provides simulated far-field mixtures for training and testing.
We emphasize that, for each real mixture in the training set, in total it has $7$ channels (i.e., $1$ close-talk plus $6$ far-field microphones), while, for each simulated training mixture, it is far-field, $6$-channel, and does not have the paired close-talk mixture.

The sampling rate is $16$ kHz.

\subsection{Synchronization of Close-talk and Far-field Mixtures}\label{sync_ct_ff}

In CHiME-4, we observe that the far-field mixtures are reasonably synchronized with each other, while significant synchronization errors exist between close-talk and far-field mixtures.
For some utterances, the time delay between the close-talk and far-field mixtures can be as large as $0.05$ seconds, which, if correct, means that the distance between the close-talk microphone and far-field array can be as large as $\sim$$17$ meters (i.e., $0.05 \times 340$, assuming that the speed of sound in the air is $340$ meters per second).
This clearly does not make sense in the CHiME-4 setup, as the speaker hand-holds the tablet while talking.
The root cause of the synchronization errors is that the close-talk microphone and far-field array in CHiME-4 are placed on, and processed by, two different devices.

If the synchronization errors between the close-talk and far-field mixtures are not properly addressed before using them to train ctPuLSEnet, the loss functions in (\ref{loss_PuLSE}) would be less effective, since the hypothesized filter taps of $\mathbf{g}_p(f)$ in (\ref{fcp_proj_mixture}) or $h_p$ in (\ref{WF_filtering}), which are hyper-parameters shared by all the training utterances, may not be able to compensate the synchronization errors for every training utterance.
To deal with this, we could use a very long linear filter to cover the maximum synchronization error of all the training mixtures.
This is however problematic, as the longer the filter is, the more likely that the filter can filter any DNN estimate (even if the estimate is a random signal) to accurately fit the estimated close-talk speech.

To mitigate the synchronization errors, we slightly modify the classic GCC-PHAT algorithm \cite{DiBiase2001, Wang2018hLocalization} to approximately align the close-talk mixture to far-field mixtures.
We emphasize that this is a pre-processing step at the very beginning of our proposed system (i.e., before training).

In detail, for each of the close-talk and far-field mixtures, we consider its magnitude in each frequency as a one-dimensional sequence, and find an integer frame shift $\hat{d}\in\ZZ$ (shared by all the frequencies) that can, at every frequency, best align the magnitude sequence of the close-talk mixture to those of the far-field mixtures.
Specifically, let $M_{p}(f)=|Y_p(\cdot,f)|\in\RR^T$ denote the magnitude sequence at frequency $f$ for microphone $p\in \{0,1,\dots,P\}$, and $R_{p}(f)=\text{FFT}\big(M_{p}(f)\big)\in\CC^T$ denote the FFT coefficients after applying a $T$-point faster Fourier transform (FFT) to the magnitude sequence.
At each frequency $f$, we first compute the GCC-PHAT coefficients between the magnitude sequence of the close-talk microphone and that of a far-field microphone $p$ by
\begin{align}
\text{GCC-}&\text{PHAT}_{p}(t,f,d) \nonumber \\
&= \mathcal{R}\Big(\frac{ R_{0}(t,f) \times R_{p}(t,f)^{*} }{ | R_{0}(t,f) | \times | R_{p}(t,f)^{*} | } e^{-j\times 2 \pi \frac{t}{T} d}\Big) \nonumber\\
&= \text{cos}\Big( \angle R_{0}(t,f) - \angle R_{p}(t,f)  - 2 \pi \frac{t}{T} d \Big), \label{GCC_PHAT_coeff_sync}
\end{align}
where $j$ denotes the imaginary unit, $d\in\ZZ$ a hypothesized frame delay, and $\mathcal{R}(\cdot)$ extracts the real component.
We then enumerate a set of hypothesized frame delays and find a delay that can produce the largest summation of the GCC-PHAT coefficients at all the far-field microphones and T-F units:
\begin{align}\label{find_best_delay}
\hat{d} = \underset{d \in \Omega}{\text{argmin}} \sum_{p=1}^P \sum_{t=0}^{T-1} \sum_{f=0}^{F-1} \text{GCC-PHAT}_{p}(t,f,d),
\end{align}
where $\Omega\in\{-D,\dots,0,\dots,D\}$ is a set of candidate frame delays with $D$, a tunable hyper-parameter, denoting a hypothesized maximum delay on each side.
If the resulting best time delay $\hat{d}$ is positive, we advance the close-talk mixture by $\hat{d}$ frames (and pad zeros to the right), and delay it by $\hat{d}$ frames otherwise (and pad zeros to the left).

Notice that this algorithm is designed to compute an integer number of frame shift for each close-talk mixture.
In our experiments, for the STFT configuration of (\ref{GCC_PHAT_coeff_sync}) and (\ref{find_best_delay}), we set the window size to $16$ ms and hop size to $1$ ms (please do not confuse this STFT configuration used in the pre-processing stage with that in the subsequent DNN training stage).
A small STFT hop size is used here, as our aim is to roughly align close-talk and far-field mixtures in this pre-processing stage.

We highlight that the alignment is performed at the granularity of $1$ ms.
Although sample-level synchronization is definitely desired, it would dramatically increase the amount of computation, as the enumeration would be conducted at the granularity of samples and this computation would become intolerable when the candidate time delay (and time advance) can be as large as $0.05$ second in CHiME-4.
In addition, we may not really need accurate sample-level synchronization, since the STFT window size of our enhancement models can be as large as $32$ ms and the linear filtering in, e.g., (\ref{loss_lp}) could account for slight synchronization errors (that are sufficiently smaller than the $32$ ms window size).

\subsection{Training Setup}\label{training_setup}

For CTSEnet, it is trained based on all the $7,138\times 6$ monaural simulated mixtures.
Following \cite{Wang2024SuperM2M}, we apply an SNR augmentation technique to the simulated training mixtures of CHiME-4.
That is, during training, we optionally modify the SNR of each simulated mixture, on the fly, by $u$ dB, with $u$ uniformly sampled from the range $[-10,+15]$ dB.
No other data augmentation is used.

For ctPuLSEnet, we use all the $7,138$ simulated and $1,600$ real utterances for training.
For monaural ctPuLSEnet, we train it on all the $(7,138+1,600)\times6$ monaural mixtures.
For $2$-channel ctPuLSEnet, at each training step we sample two microphones from the front five microphones as input, and ctPuLSEnet is trained to predict the target speech at the first of the two selected microphones.
For $6$-channel ctPuLSEnet, it stacks all the six microphones in a fixed order as input to predict the target speech at the fifth microphone.
We apply the same SNR augmentation used in CTSEnet to the simulated training mixtures when training ctPuLSEnet.
We always train ctPuLSEnet on the combination of the simulated and real mixtures using the co-learning algorithm introduced in Section \ref{PuLSE_colearning}, as there are only $1,600$ real mixtures (which amount to only $\sim$$2.7$ hours) in the training set of CHiME-4

For simplicity, we did not filter out microphone signals with any microphone failures in training and inference.
We expect ctPuLSEnet to learn to robustly deal with the failures, as it is trained on real mixtures.

\begin{figure*}
  \centering  
  \includegraphics[width=16.5cm]{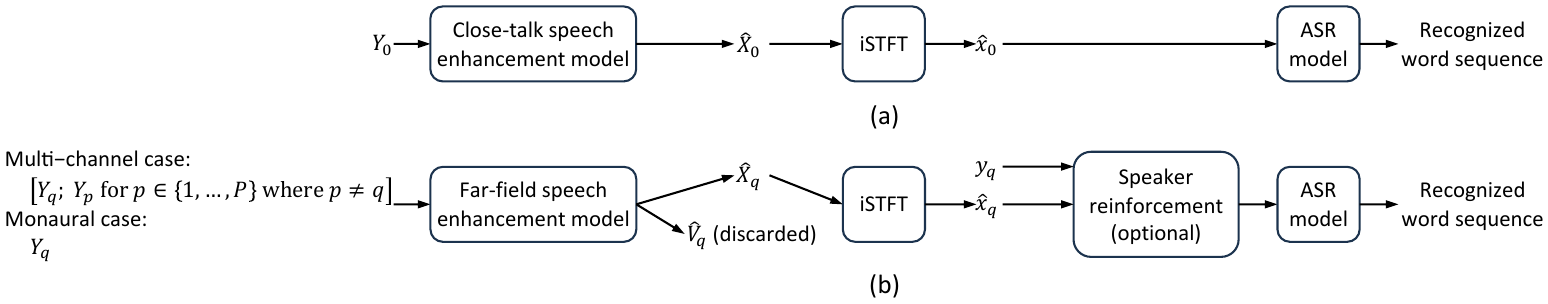}
  \caption{
    Pipeline of ASR evaluation for (a) close-talk speech enhancement; and (b) far-field speech enhancement.
    Enhanced speech (e.g., $\hat{x}_0 = \text{iSTFT}(\hat{X}_0)$ in close-talk speech enhancement) is directly fed to a pre-trained backend ASR system for recognition.
    No joint training between ASR and enhancement models is performed.
    In (b), an optional speaker reinforcement module \cite{Zorila2022SpeakerReinforcement}, which adds a scaled version of input mixture $y_q$ to $\hat{s}_q$, can be included.
  }
  \label{enh_asr_pipeline_figure}
\end{figure*}

\subsection{Miscellaneous System Configurations}
\label{configuration}

For STFT, the window and hop sizes are respectively $32$ and $8$ ms, and the square root of Hann window is used as the analysis window.

TF-GridNet \cite{Wang2022GridNetjournal} is employed as the DNN architecture.
It has shown strong separation and enhancement performance recently in serveral representitive supervised speech separation benchmarks.
Following the symbols defined in Table I of \cite{Wang2022GridNetjournal}, we set its hyper-parameters to $D=128$, $B=4$, $I=1$, $J=1$, $H=200$, $L=4$ and $E=4$ (please do not confuse the symbols with the ones defined in this paper).
The model has $\sim$$5.4$ million trainable parameters.
This configuration of TF-GridNet is used for both CTSEnet and ctPuLSEnet.

We train both CTSEnet and ctPuLSEnet on eight-second segments using a mini-batch size of one.
Adam is used as the optimizer.
The learning rate starts from $0.001$ and is halved if the loss is not improved in two epochs.

\subsection{ASR Model}

For both close-talk and far-field speech enhancement, we check whether they can result in better ASR performance by directly feeding their enhanced speech to a pre-trained ASR model for decoding, following the evaluation pipeline shown in Fig. \ref{enh_asr_pipeline_figure}(a) and (b).

The ASR model is pre-trained in a multi-conditional way on the official CHiME-4 simulated and real mixtures plus the clean speech signals in WSJ$0$ by using the script proposed in \cite{Chang2022E2EIntegration}\footnote{\url{https://github.com/espnet/espnet/blob/master/egs2/chime4/asr1/conf/tuning/train_asr_transformer_wavlm_lr1e-3_specaug_accum1_preenc128_warmup20k.yaml}}, which is available in the ESPnet toolkit.
It is an encoder-decoder- and transformer-based system, trained on WavLM features \cite{Chen2022WavLM} and using a transformer-based language model for decoding.
From the results reported in \cite{Chang2022E2EIntegration} and \cite{Masuyama2023SE}, this model is the current best ASR model on CHiME-4.

We have successfully reproduced the ASR system in \cite{Chang2022E2EIntegration}.
The mixture ASR results (shown later in row $0$ of Table \ref{results_ctPuLSEnet_1ch}) are very close to the ones reported in row $7$ of Table $1$ of \cite{Chang2022E2EIntegration}.

\subsection{Evaluation Metrics}\label{metrics}

For real-recorded mixtures, where the corresponding clean speech is unavailable for evaluation, we use word error rates (WER) as the major evaluation metric.
WER can partially reflect the intelligibility of enhanced speech.
Besides WER, DNSMOS \cite{K.A.Reddy2021DNSMOS} is employed to evaluate the quality of enhanced speech.

For simulated mixtures, where the clean speech is available, our evaluation metrics include short-time objective intelligibility (STOI) \cite{H.Taal2011}, wide-band perceptual evaluation of speech quality (wPESQ) \cite{Rix2001}, signal-to-distortion ratio (SDR) \cite{LeRoux2019}, and scale-invariant SDR (SISDR) \cite{Vincent2006a}.
They are designed to evaluate the intelligibility, quality, and accuracy of the magnitude and phase of enhanced speech. 
They are widely-adopted metrics in speech enhancement.

\subsection{Baselines}\label{baseline}

A major baseline is purely-supervised speech enhancement models trained only on the CHiME-4 simulated data.
That is, we only use Fig. \ref{colearning_figure}(b) for model training, and use exactly the same training setup as that in ctPuLSEnet.
On the other hand, since CHiME-4 is a popular public dataset, we can compare our ASR and enhancement results with the ones obtained by many earlier studies.

\subsection{Tricks to Improve ASR Performance}\label{speaker_reinforcement}

At run time, in default we feed $\hat{x}_q$ produced by ctPuLSEnet for ASR decoding.
Alternatively, following \cite{Wang2024SuperM2M} we apply an existing technique named speaker reinforcement \cite{Zorila2022SpeakerReinforcement} to mitigate speech distortion incurred by enhancement (see Fig. \ref{enh_asr_pipeline_figure}(b) for an illustration).
It adds a scaled version of the mixture signal $y_q$ back to enhanced speech $\hat{x}_q$ before performing ASR decoding.
That is, the signal sent for ASR is $\hat{x}_q + \eta \times y_q$, where $\eta$ is computed such that $10\times \text{log}_{10} \big( \| \hat{x}_q \|_2^2 / \|  \eta \times y_q \|_2^2 \big) = \gamma$ dB.
This technique has been found effective at improving ASR performance in \cite{Zorila2022SpeakerReinforcement,Wang2024SuperM2M}. 

\section{Evaluation Results}\label{evaluation_results}

\subsection{Results of Close-Talk Speech Enhancement}

Table \ref{results_close_talk} presents the results of monaural CTSEnet.

Row $0$ reports the results of unprocessed close-talk mixtures.
We observe that the ASR results are very good (e.g., $1.49\%$ on the real test set), indicating that the close-talk mixtures already have very high input SNRs and that microphone failures in the close-talk mixtures are minimal.
In comparison, the DNSMOS scores are not good (e.g., $2.50$ on the real test set), indicating that non-speech signals captured along with close-talk speech are still significant.
These non-speech signals need to be removed in order to derive high-quality pseudo-labels for far-field mixtures.

\begin{table*}[t]
\scriptsize
\centering
\sisetup{table-format=2.2,round-mode=places,round-precision=2,table-number-alignment = center,detect-weight=true,detect-inline-weight=math}
\caption{Results of CTSEnet on CHiME-4 close-talk mixtures}
\label{results_close_talk}
\setlength{\tabcolsep}{5pt}
\begin{tabular}{
c %
c %
c %
S[table-format=1.2,round-precision=2] %
S[table-format=1.2,round-precision=2] %
S[table-format=2.2,round-precision=2] %
S[table-format=2.2,round-precision=2] %
}
\toprule
& & & \multicolumn{2}{c}{DNSMOS OVRL$\uparrow$} & \multicolumn{2}{c}{WER (\%)$\downarrow$} \\
\cmidrule(lr){4-5} \cmidrule(lr){6-7}
&  & & {Val.} & {Test} & {Val.} & {Test} \\
\cmidrule(lr){4-4} \cmidrule(lr){5-5} \cmidrule(lr){6-6} \cmidrule(lr){7-7}
{Row} & {Training data} & {\multirow{1}{*}{Systems}} & {REAL} & {REAL} & {REAL} & {REAL} \\

\midrule

0 & - & Close-talk mixture & 2.818349123916309 & 2.5027792094887444 & 1.143109996 & 1.490027558 \\

\midrule

1 & S & CTSEnet & 3.2638522792785554 & 3.216589413447323 & 1.150484899885689 & 1.5554206174973143 \\

\bottomrule
\end{tabular}
\end{table*}

Row $1$ reports the results of CTSEnet trained via supervised learning on the CHiME-4 simulated mixtures (denoted as \textbf{S} in the ``Training data'' column).
Compared with row $0$, the DNSMOS scores are clearly improved (e.g., from $2.50$ to $3.22$ on the real test set).
\ZQHL{The ASR performance becomes worse, possibly because of two reasons: (a) CTSEnet is trained only on simulated data, which usually mismatches real data; and (b) CTSEnet could introduce some distortion to target speech while enhancing the mixture.}
Nonetheless, the ASR performance degrades rather slightly (e.g., from $1.49$\% to $1.56$\% WER on the real test set), meaning that the estimated close-talk speech is still of high quality and could be a good pseudo-label for far-field mixtures.

With a grain of salt, the ASR results in row $1$ can be viewed as the upper-bound performance of ctPuLSEnet.

\subsection{Key Results of Far-Field Speech Enhancement}

Table \ref{results_ctPuLSEnet_1ch} and \ref{results_ctPuLSEnet_6ch} respectively present the results of monaural and six-channel ctPuLSEnet on CHiME-4.

Let us first provide the hyper-parameter configurations of ctPuLSEnet.
The frequency-domain linear filter in (\ref{loss_lp}) is tuned to $1$-tap (i.e., in the text below (\ref{loss_lp}), $I=1$ and $J=0$).
In (\ref{WF_filtering}), the time-domain filter tap $K$ is tuned based on the set of $\{256, 128, 64, 32, 16\}$.
In this case, the linear filter length $K+1+K\in\{513, 257, 129, 65, 33\}$ is comparable to or shorter than the STFT window length, which is $512$ samples long in this paper (see the STFT configurations in Section \ref{configuration}).
The weighting term $\alpha$ in (\ref{final_loss_simu_real}) is tuned to $5$.
In both tables, when the filter tap for time-domain linear filtering (i.e., the ``$K$'' column) is denoted as ``-'', it means that frequency-domain linear filtering in (\ref{loss_lp}) is used.
Otherwise, time-domain linear filtering in (\ref{loss_lp_time}) is used.

\begin{table*}[t]
\scriptsize
\centering
\sisetup{table-format=2.2,round-mode=places,round-precision=2,table-number-alignment = center,detect-weight=true,detect-inline-weight=math}
\caption{ctPuLSE vs. purely-supervised models on CHiME-4 far-field mixtures (single-channel input)}
\label{results_ctPuLSEnet_1ch}
\resizebox{2.1\columnwidth}{!}{
\begin{tabular}{
l %
c %
c %
c %
c %
c %
c %
S[table-format=2.1,round-precision=1] %
S[table-format=2.1,round-precision=1] %
S[table-format=1.2,round-precision=2] %
S[table-format=1.3,round-precision=3] %
S[table-format=1.2,round-precision=2] %
S[table-format=1.2,round-precision=2] %
S[table-format=2.2,round-precision=2] %
S[table-format=2.2,round-precision=2] %
S[table-format=2.2,round-precision=2] %
S[table-format=2.2,round-precision=2] %
}
\toprule
& & & & & & & \multicolumn{4}{c}{SIMU Test Set (CH5)} & \multicolumn{2}{c}{DNSMOS OVRL$\uparrow$} & \multicolumn{4}{c}{WER (\%)$\downarrow$} \\

\cmidrule(lr){8-11}\cmidrule(lr){12-13}\cmidrule(lr){14-17}

& & \multirow{2}{*}[-6pt]{\begin{tabular}[c]{@{}c@{}}Training\\data\end{tabular}} & & & & &{SISDR} & {SDR} & & & 
{Val.} & {Test} & \multicolumn{2}{c}{Val.} & \multicolumn{2}{c}{Test} \\

\cmidrule(lr){12-12}\cmidrule(lr){13-13}\cmidrule(lr){14-15}\cmidrule(lr){16-17}

Row & {\multirow{1}{*}{Systems}} & & DNN arch. & $\mathcal{L}_q^{\text{simu}}$ & $\mathcal{L}_q^{\text{real}}$ & $K$ &{(dB)$\uparrow$} & {(dB)$\uparrow$} & {wPESQ$\uparrow$} & {STOI$\uparrow$} & {REAL} & {REAL} & {SIMU} & {REAL} & {SIMU} & {REAL} \\

\midrule

0 & Mixture & - & - & - & - & {-} & 7.506217692752905 & 7.539180193284881 & 1.273397988803459 & 0.8702431371034524 & 1.5207437948951077 & 1.3926635820452578 & 5.932890855 & 4.070946568 & 8.288195741 & 4.470082675 \\

\midrule

1a & Supervised & S & TF-GridNet & $\mathcal{L}_{X,q}^{\text{simu}} + \mathcal{L}_{V,q}^{\text{simu}} + \mathcal{L}_{Y,q}$ & - & {-} & \bfseries 17.295841213880163 & \bfseries 17.6653338175299 & 2.361246844584292 & 0.9597580577820082 & 3.327439949427817 & 3.20725798312422 & 3.488200589970501 & 2.1829713485010508 & 7.634478894284646 & 5.2361156522957635 \\
1b & Supervised & S & TF-GridNet & $\mathcal{L}_{X,q}^{\text{simu}} + \mathcal{L}_{V,q}^{\text{simu}}$ & - & {-} & 17.229124567725442 & 17.647178119958088 & \bfseries 2.489149991310004 & 0.9620186272022787 & 3.3173350513477446 & 3.2112040037754364 & 3.392330383480826 & 2.08709760684391 & 7.410347403810236 & 4.442057078798636 \\
1c & Supervised & S & TF-GridNet & $\mathcal{L}_{X,q}^{\text{simu}}$ & - & {-} & \bfseries 17.264941603848428 & 17.62668940438264 & 2.432730621280092 & \bfseries 0.963050892206543 & 3.3048315624147224 & 3.1872318833953543 & 3.366519174041298 & 2.238283122534017 & 7.209562943593575 & 5.166051660516605 \\

\midrule

2 & Supervised & S & \makecell{iNeuBe\\\cite{Wang2020chime}} & - & - & {-} & 15.1 & {-} & {-} & 0.954 & {-} & {-} & {-} & {-} & {-} & {-} \\

\midrule

3a & ctPuLSE & S+R & TF-GridNet & $\mathcal{L}_{X,q}^{\text{simu}} + \mathcal{L}_{V,q}^{\text{simu}} + \mathcal{L}_{Y,q}$ & $\mathcal{L}_{X,q}^{\text{real}} + \mathcal{L}_{Y,q}$ & {-} & 17.136097766684763 & 17.561826082102698 & 2.4110837602254116 & 0.9616210193433703 & 3.3567135503616337 & 3.2276916247934677 & \bfseries 3.1600294985250736 & 1.998598768391165 & \bfseries 6.840679865521106 & 3.227614554626559 \\
3b & ctPuLSE & S+R & TF-GridNet & $\mathcal{L}_{X,q}^{\text{simu}} + \mathcal{L}_{V,q}^{\text{simu}}$ & $\mathcal{L}_{X,q}^{\text{real}}$ & {-} & 16.710767206369024 & 17.254252569589138 & 2.412104403972626 & 0.9601425287251298 & 3.322187319009847 & 3.1623852528402465 & 3.558259587020649 & 2.0834101552417123 & 6.854688083675757 & 3.563921715166519 \\
3c & ctPuLSE & S+R & TF-GridNet & $\mathcal{L}_{X,q}^{\text{simu}}$ & $\mathcal{L}_{X,q}^{\text{real}}$ & {-} & 16.834305574496586 & 17.25144197396907 & 2.3900751664783018 & 0.9585712505125487 & 3.3471242118003808 & 3.1976086150428693 & 3.716814159292035 & 2.0723478004351193 & 7.639148300336197 & 3.778784623289271 \\

\midrule

4a & ctPuLSE & S+R & TF-GridNet & $\mathcal{L}_{X,q}^{\text{simu}} + \mathcal{L}_{V,q}^{\text{simu}} + \mathcal{L}_{Y,q}$ & $\mathcal{L}_{X,q}^{\text{real}} + \mathcal{L}_{Y,q}$ & $256$ & 16.71246492049911 & 17.167910347776175 & 2.293470420349728 & 0.9577850156099695 & 3.33232709343567 & 3.190962144995634 & 3.21165191740413 & 1.9801615103801764 & 7.064811355995518 & 3.40978093325237 \\
4b & ctPuLSE & S+R & TF-GridNet & $\mathcal{L}_{X,q}^{\text{simu}} + \mathcal{L}_{V,q}^{\text{simu}} + \mathcal{L}_{Y,q}$ & $\mathcal{L}_{X,q}^{\text{real}} + \mathcal{L}_{Y,q}$ & $128$ & 16.52350974308722 & 17.081067725176986 & 2.3714223765062563 & 0.9597433300866075 & 3.354864545712026 & 3.2246840220939594 & 3.4070796460176994 & 1.87691286551864 & 7.083488980201719 & 3.199588957914896 \\
4c & ctPuLSE & S+R & TF-GridNet & $\mathcal{L}_{X,q}^{\text{simu}} + \mathcal{L}_{V,q}^{\text{simu}} + \mathcal{L}_{Y,q}$ & $\mathcal{L}_{X,q}^{\text{real}} + \mathcal{L}_{Y,q}$ & $64$ & 17.016686683125567 & 17.537237975786343 & 2.4027770455136443 & 0.9612991807749263 & \bfseries 3.380592781113339 & \bfseries 3.2728282774269037 & 3.289085545722714 & \bfseries 1.8695379623142445 & 6.966753828912962 & 3.134195898921015 \\
4d & ctPuLSE & S+R & TF-GridNet & $\mathcal{L}_{X,q}^{\text{simu}} + \mathcal{L}_{V,q}^{\text{simu}} + \mathcal{L}_{Y,q}$ & $\mathcal{L}_{X,q}^{\text{real}} + \mathcal{L}_{Y,q}$ & $32$ & 16.47889514989925 & 17.141230481170066 & 2.360100678873785 & 0.9611001105517262 & 3.3450909927936032 & 3.224999633733256 & 3.3849557522123895 & 1.9875364135845716 & 6.957415016809862 & 3.54523798402541 \\
4e & ctPuLSE & S+R & TF-GridNet & $\mathcal{L}_{X,q}^{\text{simu}} + \mathcal{L}_{V,q}^{\text{simu}} + \mathcal{L}_{Y,q}$ & $\mathcal{L}_{X,q}^{\text{real}} + \mathcal{L}_{Y,q}$ & $16$ & 16.078147192231633 & 16.928411923202482 & 2.364959328012033 & 0.9594431979772853 & 3.3618951355996973 & 3.244502963142382 & 3.436578171091446 & 1.9654117039713852 & 7.228240567799776 & \bfseries 3.04077724321547 \\

\bottomrule
\end{tabular}
}
\end{table*}

\begin{table*}[t]
\scriptsize
\centering
\sisetup{table-format=2.2,round-mode=places,round-precision=2,table-number-alignment = center,detect-weight=true,detect-inline-weight=math}
\caption{ctPuLSE vs. purely-supervised models on CHiME-4 far-field mixtures (six-channel input)}
\label{results_ctPuLSEnet_6ch}
\resizebox{2.1\columnwidth}{!}{
\begin{tabular}{
l %
c %
c %
c %
c %
c %
c %
S[table-format=2.1,round-precision=1] %
S[table-format=2.1,round-precision=1] %
S[table-format=1.2,round-precision=2] %
S[table-format=1.3,round-precision=3] %
S[table-format=1.2,round-precision=2] %
S[table-format=1.2,round-precision=2] %
S[table-format=2.2,round-precision=2] %
S[table-format=2.2,round-precision=2] %
S[table-format=2.2,round-precision=2] %
S[table-format=2.2,round-precision=2] %
}
\toprule
& & & & & & & \multicolumn{4}{c}{SIMU Test Set (CH5)} & \multicolumn{2}{c}{DNSMOS OVRL$\uparrow$} & \multicolumn{4}{c}{WER (\%)$\downarrow$} \\

\cmidrule(lr){8-11}\cmidrule(lr){12-13}\cmidrule(lr){14-17}

& & \multirow{2}{*}[-6pt]{\begin{tabular}[c]{@{}c@{}}Training\\data\end{tabular}} & & & & & {SISDR} & {SDR} & & & 
{Val.} & {Test} & \multicolumn{2}{c}{Val.} & \multicolumn{2}{c}{Test} \\

\cmidrule(lr){12-12}\cmidrule(lr){13-13}\cmidrule(lr){14-15}\cmidrule(lr){16-17}

{\multirow{1}{*}[0pt]{\rotatebox[origin=c]{0}{Row}}} & {\multirow{1}{*}{Systems}} & & DNN arch. & $\mathcal{L}_q^{\text{simu}}$ & $\mathcal{L}_q^{\text{real}}$ & $K$ & {(dB)$\uparrow$} & {(dB)$\uparrow$} & {wPESQ$\uparrow$} & STOI$\uparrow$ & {REAL} & {REAL} & {SIMU} & {REAL} & {SIMU} & {REAL} \\

\midrule

0 & Mixture & - & - & - & - & {-} & 7.506217692752905 & 7.539180193284881 & 1.273397988803459 & 0.8702431371034524 & 1.5207437948951077 & 1.3926635820452578 & 5.932890855 & 4.070946568 & 8.288195741 & 4.470082675 \\

\midrule

1a & Supervised & S & TF-GridNet & $\mathcal{L}_{X,q}^{\text{simu}} + \mathcal{L}_{V,q}^{\text{simu}} + \mathcal{L}_{Y,q}$ & - & {-} & 22.91391611189553 & \bfseries 23.32200724741185 & \bfseries 3.344010585185253 & 0.987202710663977 & 2.1200781315271033 & 1.776820183864461 & 0.8628318584070796 & 23.164570965006084 & 1.3401195367949197 & 46.713998785557476 \\
1b & Supervised & S & TF-GridNet & $\mathcal{L}_{X,q}^{\text{simu}} + \mathcal{L}_{V,q}^{\text{simu}}$ & - & {-} & \bfseries 22.9554724250779 & \bfseries 23.339479570583496 & 3.289384816090266 & \bfseries 0.9875730485728832 & 2.0635060826064993 & 1.6312558424754935 & 0.855457227138643 & 40.27065894760131 & 1.3167725065371685 & 64.7111028072306 \\
1c & Supervised & S & TF-GridNet & $\mathcal{L}_{X,q}^{\text{simu}}$ & - & {-} & 22.809102938030705 & 23.20391760625078 & 3.2152784886685284 & 0.9872708922673862 & 2.0432281144655318 & 1.6272474622321742 & \bfseries 0.8259587020648967 & 53.54179726391091 & \bfseries 1.2980948823309675 & 74.21645102526975 \\

\midrule

2a & Supervised & S & \makecell{iNeuBe\\\cite{Wang2020chime}} & - & - & {-} & 22.0 & 22.4 & {-} & 0.986 & {-} & {-} & {-} & {-} & {-} & {-} \\
2b & Supervised & S & \makecell{SpatialNet\\\cite{Quan2024SpatialNet}} & - & - & {-} & 22.1 & 22.3 & 2.88 & 0.983 & {-} & {-} & {-} & {-} & {-} & {-} \\
2c & Supervised & S & \makecell{USES\\\cite{Zhang2023USES}} & - & - & {-} & {-} & 20.6 & 3.16 & 0.983 & {-} & {-} & {-} & {-} & 4.2 & 78.1 \\
2d & Supervised & S & \makecell{USES2\\\cite{Zhang2024USES2}} & - & - & {-} & {-} & 18.8 & 2.94 & 0.979 & {-} & 2.96 & {-} & {-} & 4.6 & 12.1 \\

\midrule

3a & ctPuLSE & S+R & TF-GridNet & $\mathcal{L}_{X,q}^{\text{simu}} + \mathcal{L}_{V,q}^{\text{simu}} + \mathcal{L}_{Y,q}$ & $\mathcal{L}_{X,q}^{\text{real}} + \mathcal{L}_{Y,q}$ & {-} & 22.551012031959765 & 22.84897036980553 & 3.114229646144491 & 0.9849237553000239 & 3.384385940797479 & 3.2789563173090364 &  0.8185840707964602 & 1.283233157564807 & 1.3914830033619723 & 1.653510205988136 \\
3b & ctPuLSE & S+R & TF-GridNet & $\mathcal{L}_{X,q}^{\text{simu}} + \mathcal{L}_{V,q}^{\text{simu}}$ & $\mathcal{L}_{X,q}^{\text{real}}$ & {-} & 22.805953327453498 & \bfseries 23.27448130848245 & 3.262422973278797 & 0.9869820737165041 & \bfseries 3.38914009797696 & 3.2816045025459695 & 0.8702064896755162 & 1.2721708027582138 & 1.3494583488980201 & 1.7329160633378485 \\
3c & ctPuLSE & S+R & TF-GridNet & $\mathcal{L}_{X,q}^{\text{simu}}$ & $\mathcal{L}_{X,q}^{\text{real}}$ & {-} & 22.795234206047926 & 23.116705751770436 & 3.2897390328573457 & 0.9869850450991228 & \bfseries 3.394377605833281 & \bfseries 3.3028374018258155 & 0.8443952802359883 & 1.2869206091670048 & \bfseries 1.2980948823309675 & \bfseries 1.5647624830678688 \\

\midrule

4a & ctPuLSE & S+R & TF-GridNet & $\mathcal{L}_{X,q}^{\text{simu}} + \mathcal{L}_{V,q}^{\text{simu}} + \mathcal{L}_{Y,q}$ & $\mathcal{L}_{X,q}^{\text{real}} + \mathcal{L}_{Y,q}$ & $256$ & 22.646991805596784 & 22.958123369688533 & 3.112808776714585 & 0.9855840736812684 & 3.3676542193702486 & 3.2494355439977562 & 0.855457227138643 & 1.3053578671779933 & 1.3401195367949197 & 1.7702835256200665 \\
4b & ctPuLSE & S+R & TF-GridNet & $\mathcal{L}_{X,q}^{\text{simu}} + \mathcal{L}_{V,q}^{\text{simu}} + \mathcal{L}_{Y,q}$ & $\mathcal{L}_{X,q}^{\text{real}} + \mathcal{L}_{Y,q}$ & $128$ & 22.877807343547996 & 23.22293037669213 & 3.2195719266479665 & 0.9861512576974908 & 3.340576934824889 & 3.251119989958086 & 0.8370206489675517 & 1.323795125188982 & 1.3354501307433694 & 1.700219533840908 \\
4c & ctPuLSE & S+R & TF-GridNet & $\mathcal{L}_{X,q}^{\text{simu}} + \mathcal{L}_{V,q}^{\text{simu}} + \mathcal{L}_{Y,q}$ & $\mathcal{L}_{X,q}^{\text{real}} + \mathcal{L}_{Y,q}$ & $64$ & 22.804458647966385 & 23.199798023054065 & 3.199864766453252 & 0.9861371389833652 & \bfseries 3.3853782118151963 & 3.289810700044853 & 0.8591445427728614 & \bfseries 1.2537335447472252 & 1.3681359731042212 & 1.700219533840908 \\
4d & ctPuLSE & S+R & TF-GridNet & $\mathcal{L}_{X,q}^{\text{simu}} + \mathcal{L}_{V,q}^{\text{simu}} + \mathcal{L}_{Y,q}$ & $\mathcal{L}_{X,q}^{\text{real}} + \mathcal{L}_{Y,q}$ & $32$ & 22.71326851031997 & 23.082258429569077 & 3.2345488795728397 & 0.9860629056563212 & \bfseries 3.391553352167839 & \bfseries 3.3023168560798366 & 0.8702064896755162 & 1.2795457059626092 & 1.307433694434068 & 1.6721939371292446 \\
4e & ctPuLSE & S+R & TF-GridNet & $\mathcal{L}_{X,q}^{\text{simu}} + \mathcal{L}_{V,q}^{\text{simu}} + \mathcal{L}_{Y,q}$ & $\mathcal{L}_{X,q}^{\text{real}} + \mathcal{L}_{Y,q}$ & $16$ & 22.581081933595918 & 22.95786733754969 & 3.24542808776552 & 0.985162009465439 & 3.3795466465731185 & 3.273793347574731 & 0.8886430678466077 & 1.2611084479516206 & 1.4194994396712738 & 1.7983091223317295 \\

\bottomrule
\end{tabular}
}
\end{table*}

The key results of this paper are presented in row $0$, $1$a and $3$a of Table \ref{results_ctPuLSEnet_1ch} and \ref{results_ctPuLSEnet_6ch}.
Comparing row $1$a with $0$, we observe that, on the simulated test data, purely-supervised learning based TF-GridNet trained on simulated data obtains strong enhancement results\footnote{By ``enhancement results'', we mean SISDR, SDR, wPESQ and STOI scores.} (e.g., $17.3$ vs. $7.5$ dB SISDR in Table \ref{results_ctPuLSEnet_1ch}, and $22.9$ vs. $7.5$ dB SISDR in Table \ref{results_ctPuLSEnet_6ch}), and strong ASR performance (e.g., $7.63$\% vs. $8.29$\% WER in Table \ref{results_ctPuLSEnet_1ch}, and $1.34$\% vs. $8.29$\% WER in Table \ref{results_ctPuLSEnet_6ch}).
In addition, the enhancement results on the simulated test data obtained by TF-GridNet in row $1$a are better than strong existing models such as iNeuBe \cite{Wang2020chime, Lu2022L3DAS}, SpatialNet \cite{Quan2024SpatialNet}, USES \cite{Zhang2023USES} and USES2 \cite{Zhang2024USES2}.
However, the performance on the real test data is limited.
For example, the ASR performance on the real test data is degraded compared to just using unprocessed mixtures for ASR decoding (e.g., $4.47$\% vs. $5.24$\% WER in Table \ref{results_ctPuLSEnet_1ch}).
This degradation is much more severe in multi-channel cases (e.g., $4.47$\% vs. $46.71$\% WER in Table \ref{results_ctPuLSEnet_6ch}), possibly because simulated inter-microphone characteristics are more likely to mismatch those in real-recorded multi-channel data, compared with monaural cases where only one microphone needs to be simulated.
These problems are widely-observed in earlier robust ASR studies \cite{HaebUmbach2019SPM,Haeb-Umbach2020}, largely because (a) on real data, enhancement models trained on simulated data usually introduce speech distortion detrimental to ASR; and (b) simulated training data is often mismatched with real-recorded test data.
In row $3$a, our proposed ctPuLSEnet, trained on simulated and real data combined (denoted as \textbf{S+R} in the ``Training data'' column), obtains clearly better ASR performance on the real test set over row $1$a and $0$ (e.g., $3.23$\% vs. $5.24$\% and $4.47$\% in Table \ref{results_ctPuLSEnet_1ch}, and $1.65$\% vs. $46.71$\% and $4.47$\% in Table \ref{results_ctPuLSEnet_6ch}).
These results indicate that ctPuLSE is an effective mechanism for learning from real-recorded data, and can yield enhancement models with better generalizability to real data than purely-supervised approaches which train enhancement models only on simulated data.

\subsection{Ablation Results of Far-Field Speech Enhancement}

Next, we present several ablation results of ctPuLSEnet.

Comparing row $3$a with $3$b and $3$c in Table \ref{results_ctPuLSEnet_1ch}, we observe that configuring ctPuLSEnet to estimate noise besides target speech and at the same time including both the loss on the noise estimate and the mixture-constraint loss (i.e., row $3$a) lead to better enhancement and ASR performance.
Comparing $3$a with $1$a, $3$b with $1$b, and $3$c with $1$c, we observe that ctPuLSE obtains better ASR performance on the real test data for the various loss functions used for the simulated data. 

In $4$a-$4$e of Table \ref{results_ctPuLSEnet_1ch}, we switch from frequency-domain linear filtering to time-domain linear filtering, and experiment with various filter lengths by tuning $K$ (defined in the text below (\ref{WF_filtering})) based on the set of $\{256, 128, 64, 32, 16\}$.
We observe that setting $K$ to $64$ (in row $4$c) produces the best ASR performance on the real validation set among various options, and the ASR performance is also better than row $3$a on the real validation set (i.e., $1.87$\% vs. $2.00$\% WER in Table \ref{results_ctPuLSEnet_1ch}).
Similar trend is also observed in the six-channel-input case in Table \ref{results_ctPuLSEnet_6ch}.
We therefore choose the setup in row $4$c for the rest of experiments in this paper.

\ZQHL{
\subsection{More Ablation Results of Far-Field Speech Enhancement}

This subsection provides more ablation results of ctPuLSE to show the effectiveness of some key hyper-parameters we use for training, including (a) whether pre-training the DNN in ctPuLSE on simulated mixtures; (b) tuning the batch size used for training ctPuLSE; and (c) tuning the weighting term for the loss on simulated mixtures (i.e., $\alpha$ in Eq. (\ref{final_loss_simu_real})).
The results are obtained based on the $6$-channel system configuration in row $4$c of Table \ref{results_ctPuLSEnet_6ch}.
}

\ZQHL{
\subsubsection{With or Without Pre-training on Simulated Mixtures}

In default, we train the DNN in ctPuLSE from scratch via co-learning based on simulated and real mixtures.
Another strategy is to first pre-train the DNN via supervised learning on simulated mixtures, and then fine-tune the DNN on real mixtures or on both simulated and real mixtures.
We compare their performance in row $4$c, $5$a and $5$b of Table \ref{results_ctPuLSEnet_6ch_finetuning}.
We observe that the fine-tuning approach does not produce better WER on the REAL validation set than the default approach.
}

\begin{table*}[t]
\scriptsize
\centering
\sisetup{table-format=2.2,round-mode=places,round-precision=2,table-number-alignment = center,detect-weight=true,detect-inline-weight=math}
\caption{\ZQHL{Results of ctPuLSE with and without supervised pre-training (based on simulated training mixtures) on CHiME-4 far-field mixtures (six-channel input)}}
\label{results_ctPuLSEnet_6ch_finetuning}
\setlength{\tabcolsep}{5pt}
\begin{tabular}{
c %
c %
c %
S[table-format=1.2,round-precision=2] %
S[table-format=1.2,round-precision=2] %
S[table-format=2.2,round-precision=2] %
S[table-format=2.2,round-precision=2] %
S[table-format=2.2,round-precision=2] %
S[table-format=2.2,round-precision=2] %
}
\toprule
& & & \multicolumn{2}{c}{DNSMOS OVRL$\uparrow$} & \multicolumn{4}{c}{WER (\%)$\downarrow$} \\

\cmidrule(lr){4-5}\cmidrule(lr){6-9}

& & \multirow{2}{*}[-6pt]{\begin{tabular}[c]{@{}c@{}}Training\\data\end{tabular}} & 
{Val.} & {Test} & \multicolumn{2}{c}{Val.} & \multicolumn{2}{c}{Test} \\

\cmidrule(lr){4-4}\cmidrule(lr){5-5}\cmidrule(lr){6-7}\cmidrule(lr){8-9}

{\multirow{1}{*}[0pt]{\rotatebox[origin=c]{0}{Row}}} & {\multirow{1}{*}{Systems}} & & {REAL} & {REAL} & {SIMU} & {REAL} & {SIMU} & {REAL} \\

\midrule

0 & Mixture & - & 1.5207437948951077 & 1.3926635820452578 & 5.932890855 & 4.070946568 & 8.288195741 & 4.470082675 \\

\midrule

1a of Table \ref{results_ctPuLSEnet_6ch} & Supervised & S & 2.1200781315271033 & 1.776820183864461 & 0.8628318584070796 & 23.164570965006084 & 1.3401195367949197 & 46.713998785557476 \\

\midrule

4c of Table \ref{results_ctPuLSEnet_6ch} & ctPuLSE & S+R & \bfseries 3.3853782118151963 & \bfseries 3.289810700044853 & \bfseries 0.8591445427728614 & \bfseries 1.2537335447472252 & 1.3681359731042212 & 1.700219533840908 \\

\midrule

5a & Supervised$\rightarrow$ctPuLSE & S$\rightarrow$S+R & 3.3689970322239375 & 3.2477574351902017 & 1.0803834808259588 & 1.283233157564807 & 2.0638774747852074 & \bfseries 1.658181138773413 \\
5b & Supervised$\rightarrow$ctPuLSE & S$\rightarrow$R & 3.350789040414652 & 3.2209365158718644 & 0.866519174041298 & 1.283233157564807 & \bfseries 1.2934254762794173 & 1.765612592834789 \\

\bottomrule
\end{tabular}

\vspace{0.3cm}

\scriptsize
\centering
\sisetup{table-format=2.2,round-mode=places,round-precision=2,table-number-alignment = center,detect-weight=true,detect-inline-weight=math}
\caption{\ZQHL{Results of ctPuLSE trained with different batch sizes on CHiME-4 far-field mixtures (six-channel input)}}
\label{results_ctPuLSEnet_6ch_batch_sizes}
\setlength{\tabcolsep}{5pt}
\begin{tabular}{
l %
c %
c %
S[table-format=1.2,round-precision=2] %
S[table-format=1.2,round-precision=2] %
S[table-format=2.2,round-precision=2] %
S[table-format=2.2,round-precision=2] %
S[table-format=2.2,round-precision=2] %
S[table-format=2.2,round-precision=2] %
}
\toprule
& & & \multicolumn{2}{c}{DNSMOS OVRL$\uparrow$} & \multicolumn{4}{c}{WER (\%)$\downarrow$} \\

\cmidrule(lr){4-5}\cmidrule(lr){6-9}

& & \multirow{2}{*}[-6pt]{\begin{tabular}[c]{@{}c@{}}Batch\\size\end{tabular}} & 
{Val.} & {Test} & \multicolumn{2}{c}{Val.} & \multicolumn{2}{c}{Test} \\

\cmidrule(lr){4-4}\cmidrule(lr){5-5}\cmidrule(lr){6-7}\cmidrule(lr){8-9}

{\multirow{1}{*}[0pt]{\rotatebox[origin=c]{0}{Row}}} & {\multirow{1}{*}{Systems}} & & {REAL} & {REAL} & {SIMU} & {REAL} & {SIMU} & {REAL} \\

\midrule

0 & Mixture & {-} & 1.5207437948951077 & 1.3926635820452578 & 5.932890855 & 4.070946568 & 8.288195741 & 4.470082675 \\

\midrule

4c of Table \ref{results_ctPuLSEnet_6ch} & ctPuLSE & 1 & \bfseries 3.3853782118151963 & \bfseries 3.289810700044853 & 0.8591445427728614 & \bfseries 1.2537335447472252 & 1.3681359731042212 & 1.700219533840908 \\

\midrule

6a & ctPuLSE & 2 & \bfseries 3.3860701514691605 & \bfseries 3.287134559354147 & \bfseries 0.8333333333333333 & 1.283233157564807 & 1.3681359731042212 & \bfseries 1.653510205988136 \\
6b & ctPuLSE & 4 & 3.3726170476090895 & 3.2460520165672135 & 0.892330383480826 & 1.2647958995538184 & \bfseries 1.3354501307433694 & 1.7095613994114623 \\
6c & ctPuLSE & 6 & 3.3751361641052284 & 3.2463847062269977 & 0.8812684365781712 & 1.3274825767911795 & 1.3961524094135227 & 1.7282451305525713 \\

\bottomrule
\end{tabular}
\end{table*}

\begin{table*}[t]
\scriptsize
\centering
\sisetup{table-format=2.2,round-mode=places,round-precision=2,table-number-alignment = center,detect-weight=true,detect-inline-weight=math}
\caption{\ZQHL{Results of ctPuLSE trained with different weighting term $\alpha$ (for the loss in Eq. (\ref{final_loss_simu_real})) on CHiME-4 far-field mixtures (six-channel input)}}
\label{results_ctPuLSEnet_6ch_alpha}
\setlength{\tabcolsep}{5pt}
\begin{tabular}{
l %
c %
c %
S[table-format=1.2,round-precision=2] %
S[table-format=1.2,round-precision=2] %
S[table-format=2.2,round-precision=2] %
S[table-format=2.2,round-precision=2] %
S[table-format=2.2,round-precision=2] %
S[table-format=2.2,round-precision=2] %
}
\toprule
& & & \multicolumn{2}{c}{DNSMOS OVRL$\uparrow$} & \multicolumn{4}{c}{WER (\%)$\downarrow$} \\

\cmidrule(lr){4-5}\cmidrule(lr){6-9}

& & & 
{Val.} & {Test} & \multicolumn{2}{c}{Val.} & \multicolumn{2}{c}{Test} \\

\cmidrule(lr){4-4}\cmidrule(lr){5-5}\cmidrule(lr){6-7}\cmidrule(lr){8-9}

{\multirow{1}{*}[0pt]{\rotatebox[origin=c]{0}{Row}}} & {\multirow{1}{*}{Systems}} & {$\alpha$ in Eq. (\ref{final_loss_simu_real})} & {REAL} & {REAL} & {SIMU} & {REAL} & {SIMU} & {REAL} \\

\midrule

0 & Mixture & {-} & 1.5207437948951077 & 1.3926635820452578 & 5.932890855 & 4.070946568 & 8.288195741 & 4.470082675 \\

\midrule

4c of Table \ref{results_ctPuLSEnet_6ch} & ctPuLSE & 5 & 3.3853782118151963 & 3.289810700044853 & \bfseries 0.8591445427728614 & 1.2537335447472252 & 1.3681359731042212 & 1.700219533840908 \\

\midrule

7a & ctPuLSE & 10 & 3.4014498035068943 & \bfseries 3.296591379797464 & \bfseries 0.8591445427728614 & 1.2611084479516206 & 1.363466567052671 & 1.7095613994114623 \\
7b & ctPuLSE & 2.5 & \bfseries 3.40514200564537 & \bfseries 3.302443070999696 & 0.8702064896755162 & 1.283233157564807 & \bfseries 1.307433694434068 & \bfseries 1.6068008781353636 \\
7c & ctPuLSE & 1.0 & 3.397188011452271 & 3.292664756077591 & 0.866519174041298 & 1.2537335447472252 & 1.3214419125887188 & 1.7282451305525713 \\
7d & ctPuLSE & 0.5 & 3.390984208319801 & 3.2891417328591523 &  0.8702064896755162 & \bfseries 1.2352962867362366 & 1.3821441912588718 & 1.6908776682703536 \\
7e & ctPuLSE & 0.25 & 3.388038322580972 & 3.2800156524846114 & 0.8886430678466077 & 1.268483351156016 & 1.4428464699290251 & 1.7142323321967398 \\
7f & ctPuLSE & 0.1 & 3.3724809047049504 & 3.266704502908784 & 0.9845132743362833 & 1.3606696412109591 & 1.5035487485991782 & 1.807650987902284 \\

\bottomrule
\end{tabular}
\end{table*}

\begin{table*}[t]
\scriptsize
\centering
\sisetup{table-format=2.2,round-mode=places,round-precision=2,table-number-alignment = center,detect-weight=true,detect-inline-weight=math}
\caption{Comparison of ctPuLSE with other approaches on CHiME-4 far-field mixtures}
\label{results_comparison_with_others}
\setlength{\tabcolsep}{5pt}
\begin{tabular}{
l %
c %
c %
c %
c %
S[table-format=1.2,round-precision=2] %
S[table-format=1.2,round-precision=2] %
S[table-format=2.2,round-precision=2] %
S[table-format=2.2,round-precision=2] %
S[table-format=2.2,round-precision=2] %
S[table-format=2.2,round-precision=2] %
}
\toprule
& & & & & \multicolumn{2}{c}{DNSMOS OVRL$\uparrow$} & \multicolumn{4}{c}{WER (\%)$\downarrow$} \\
\cmidrule(lr){6-7}\cmidrule(lr){8-11}
& & & \multirow{2}{*}[-5pt]{\begin{tabular}[c]{@{}c@{}}Joint\\training? \end{tabular}} & \multirow{2}{*}[-5pt]{\begin{tabular}[c]{@{}c@{}}\#input\\mics \end{tabular}} & {Val.} & {Test} & \multicolumn{2}{c}{Val.} & \multicolumn{2}{c}{Test} \\
\cmidrule(lr){6-6}\cmidrule(lr){7-7}\cmidrule(lr){8-9}\cmidrule(lr){10-11}
{\multirow{1}{*}[0pt]{\rotatebox[origin=c]{0}{Row}}} & {\multirow{1}{*}{Systems}} & Frontend & & & {REAL} & {REAL} & {SIMU} & {REAL} & {SIMU} & {REAL} \\

\midrule

0 & Mixture & - & - & 1 & 1.5207437948951077 & 1.3926635820452578 & 5.932890855 & 4.070946568 & 8.288195741 & 4.470082675 \\

\midrule

1a & IRIS \cite{Chang2022E2EIntegration} & Conv-TasNet & \xmark & 1 & {-} & {-} & 5.96 & 4.37 & 13.52 & 12.11 \\
1b & IRIS \cite{Chang2022E2EIntegration} & Conv-TasNet & \cmark & 1 & {-} & {-} & \bfseries 3.16 & 2.03 & \bfseries 6.12 & 3.92 \\
2 & SuperM2M \cite{Wang2024SuperM2M} & TF-GridNet & \xmark & 1 & 3.163384205610393 & 3.033267061570577 & 3.392330383 & \bfseries 1.836350897 & 6.569854314 & \bfseries 3.03610631 \\
3 & ctPuLSE ($4$c of Table \ref{results_ctPuLSEnet_1ch}) & TF-GridNet & \xmark & 1 & \bfseries 3.380592781113339 & \bfseries 3.2728282774269037 & 3.289085545722714 & 1.8695379623142445 & 6.966753828912962 & 3.134195898921015 \\

\midrule

4a & MultiIRIS \cite{Masuyama2023SE} & Neural WPD & \xmark & 2 & {-} & {-} & 2.28 & 2.06 & 2.30 & 3.63 \\
4b & MultiIRIS \cite{Masuyama2023SE} & Neural WPD & \cmark & 2 & {-} & {-} & 2.04 & 1.66 & \bfseries 2.04 & 2.65 \\
5 & SuperM2M \cite{Wang2024SuperM2M} & TF-GridNet & \xmark & 2 & 3.0069181637929403 & 2.903423564541131 & \bfseries 1.497050147 & \bfseries 1.401231608 & 2.082555098 & \bfseries 1.943108038 \\
6 & ctPuLSE ($4$c of Table \ref{results_ctPuLSEnet_6ch}) & TF-GridNet & \xmark & 2 & \bfseries 3.404649122969891 & \bfseries 3.3121048136906417 & 1.5744837758112095 & 1.4528559312659022 & 2.325364213672021 & 2.08323602223364 \\

\midrule

7a & MultiIRIS \cite{Masuyama2023SE} & Neural WPD & \xmark & 6 & {-} & {-} & 1.19 & 1.32 & 1.29 & 1.85 \\
7b & MultiIRIS \cite{Masuyama2023SE} & Neural WPD & \cmark & 6 & {-} & {-} & 1.22 & 1.33 & \bfseries 1.24 & 1.77 \\
8 & SuperM2M \cite{Wang2024SuperM2M} & TF-GridNet & \xmark & 6 & 2.839865817670837 & 2.7495830637026297 & \bfseries 0.829646017 & 1.257420996 & 1.368135973 & \bfseries 1.606800878 \\
9 & ctPuLSE ($4$c of Table \ref{results_ctPuLSEnet_6ch}) & TF-GridNet & \xmark & 6 & \bfseries 3.3853782118151963 & \bfseries 3.289810700044853 & 0.8591445427728614 & \bfseries 1.2537335447472252 & 1.3681359731042212 & 1.700219533840908 \\

\bottomrule
\multicolumn{11}{l}{
\textit{Note $\#1$}: The best scores are highlighted in bold in each of the $1$-, $2$- and $6$-channel tasks separately.
}\\
\multicolumn{11}{l}{
\textit{Note $\#2$}: SuperM2M and ctPuLSE use exactly the same TF-GridNet architecture.
}%
\end{tabular}
\end{table*}

\begin{table*}[t]
\scriptsize
\centering
\sisetup{table-format=2.2,round-mode=places,round-precision=2,table-number-alignment = center,detect-weight=true,detect-inline-weight=math}
\caption{Effects of speaker reinforcement}
\label{results_speaker_reinforce}
\setlength{\tabcolsep}{5pt}
\begin{tabular}{
l %
c %
c %
S[table-format=2,round-precision=0] %
S[table-format=2.2,round-precision=2] %
S[table-format=2.2,round-precision=2] %
S[table-format=2.2,round-precision=2] %
S[table-format=2.2,round-precision=2] %
}
\toprule
& & & & \multicolumn{4}{c}{WER (\%)$\downarrow$} \\
\cmidrule(lr){5-8}
& & & & \multicolumn{2}{c}{Val.} & \multicolumn{2}{c}{Test} \\
\cmidrule(lr){5-6}\cmidrule(lr){7-8}
{\multirow{1}{*}[0pt]{\rotatebox[origin=c]{0}{Row}}} & {\multirow{1}{*}{Systems}} & {\#input mics} & {Speaker reinforcement $\gamma$ (dB)} & {SIMU} & {REAL} & {SIMU} & {REAL} \\

\midrule

0 & Mixture & 1 & {-} & 5.932890855 & 4.070946568 & 8.288195741 & 4.470082675 \\

\midrule

1a & SuperM2M \cite{Wang2024SuperM2M} & 1 & {-} & 3.392330383 & 1.836350897 & 6.569854314 & 3.03610631 \\
1b & SuperM2M \cite{Wang2024SuperM2M} & 1 & 10 & 2.400442477 & \bfseries 1.640915962 & \bfseries 4.538662682 & \bfseries 2.400859451 \\
2a & ctPuLSE & 1 & {-} & 3.289085545722714 & 1.8695379623142445 & 6.966753828912962 & 3.134195898921015 \\
2b & ctPuLSE & 1 & 10 & \bfseries 2.293510324483776 & 1.66672812419337 & 4.758124766529698 & 2.550329300761362 \\

\midrule

3a & SuperM2M \cite{Wang2024SuperM2M} & 2 & {-} & 1.497050147 & 1.401231608 & 2.082555098 & 1.943108038 \\
3b & SuperM2M \cite{Wang2024SuperM2M} & 2 & 10 & \bfseries 1.28318584 & \bfseries 1.331170028 & \bfseries 1.881770638 & \bfseries 1.835676584 \\
4a & ctPuLSE & 2 & {-} & 1.5744837758112095 & 1.4528559312659022 & 2.325364213672021 & 2.08323602223364 \\
4b & ctPuLSE & 2 & 10 & 1.4454277286135693 & 1.3569821896087613 & 2.003175196115054 & 1.845018450184502 \\

\midrule

5a & SuperM2M \cite{Wang2024SuperM2M} & 6 & {-} & \bfseries 0.829646017 & 1.257420996 & \bfseries 1.368135973 & 1.606800878 \\
5b & SuperM2M \cite{Wang2024SuperM2M} & 6 & 10 & \bfseries 0.833333333 & 1.231608835 & \bfseries 1.372805379 & 1.578775281 \\
6a & ctPuLSE & 6 & {-} & 0.8591445427728614 & 1.2537335447472252 & \bfseries 1.3681359731042212 & 1.700219533840908 \\
6b & ctPuLSE & 6 & 10 & 0.855457227138643 & \bfseries 1.2242339319296434 & \bfseries 1.3728053791557714 & \bfseries 1.5647624830678688 \\

\midrule

\rowcolor{shadecolor}
7 & Close-talk Mixture & - & {-} & {-} & 1.143109996 & {-} & 1.490027558 \\

\bottomrule
\multicolumn{8}{l}{
\textit{Note}: Oracle results (i.e., directly using close-talk mixtures for ASR decoding)  are marked in grey.
}
\end{tabular}
\end{table*}

\ZQHL{
\subsubsection{Effects of Batch Size}

In default, the mini-batch size is set to one for model training.
In Table \ref{results_ctPuLSEnet_6ch_batch_sizes}, we report the results of ctPuLSE when the mini-batch size is configured larger than one.
From the results, we do not observe large difference in performance, and setting the mini-batch size to one yields the best WER on the REAL validation set.
}

\ZQHL{
\subsubsection{Weighting Term Between Losses on Simulated and Real Mixtures}

In default, this paper sets the weighting term between the losses on simulated and real mixtures (i.e., $\alpha$ in Eq. (\ref{final_loss_simu_real})) to $5$.
In Table \ref{results_ctPuLSEnet_6ch_alpha}, we report the results of enumerating $\alpha$ in the set of $\{10, 5, 2.5, 1.0, 0.5, 0.25, 0.1\}$.
From the results, we only observe slight fluctuation in performance, and setting $\alpha$ to $5$ produces strong performance among all the values.

Notice that there are $7,138$ simulated mixtures and $1,600$ real mixtures in the training set.
That is, there are many more simulated mixtures than real mixtures.
In this case, we however do not observe too much performance difference even if setting $\alpha$ to a value larger than one.
This is possibly because the loss value on simulated mixtures is much smaller than that on real mixtures, as the target signal for real mixtures is much harder to fit.
}

\subsection{Comparison with Other Approaches}

In Table \ref{results_comparison_with_others}, we compare the performance of ctPuLSEnet with IRIS \cite{Chang2022E2EIntegration}, multi-channel IRIS (MultiIRIS) \cite{Masuyama2023SE} and SuperM2M \cite{Wang2024SuperM2M} on the $1$-, $2$-, and $6$-channel tasks of CHiME-4.

IRIS and MultiIRIS were the state-of-the-art systems on CHiME-4 before SuperM2M.
Comparing row $1$a with $1$b, $4$a with $4$b, and $7$a with $7$b, we observe that IRIS and MultiIRIS need joint frontend-backend training to achieve strong ASR performance, especially in the $1$- and $2$-channel cases.

SuperM2M \cite{Wang2024SuperM2M}, even without joint frontend-backend training, achieves better ASR performance on the real data than IRIS and MultiIRIS.
However, the enhancement results (measured by DNSMOS OVRL) of SuperM2M on the real data are not strong.
Upon listening to its processed signals\footnote{See a sound demo at \url{https://zqwang7.github.io/demos/ctPuLSE_demo/index.html}, which provides a comparison between SuperM2M and ctPuLSE.}, we observe that it cannot suppress noises sufficiently and tends to maintain some noise signals in its estimate of target speech, especially in multi-channel cases.
This is reflected by the DNSMOS OVRL scores in row $2$, $5$ and $8$, and it is quite counter-intuitive that the score decreases when the number of input microphones increases (e.g., from $3.03$ in the monaural case down to $2.90$ in the $2$-channel case and down to $2.75$ in the $6$-channel case on the real test data).
We think that the mediocre enhancement score is likely because, in SuperM2M \cite{Wang2024SuperM2M}, the loss function is defined on observed mixtures (i.e., between each observed mixture and reconstructed mixture, which is obtained by summing up linearly-filtered source estimates) rather than on individual source estimates.
In this case, the DNN would only have a \textit{weak supervision} regarding what the target sources are.
That is, the cues leveraged for training the SuperM2M enhancement models are (a) there should be two sources; and (b) their estimates, after linear filtering, should add up to each mixture.
Such a supervision could be too weak for the DNN to achieve good enhancement.

In comparison, in ctPuLSE, the loss function includes a loss term on the pseudo-target speech provided by CTSEnet (i.e., $\mathcal{L}_{X}^{\text{real}}$ in (\ref{loss_PuLSE})).
As long as the pseudo-target is reasonably accurate, ctPuLSE is expected to more accurately suppress non-target signals.
Comparing row $3$ with $2$, $6$ with $5$, and $9$ with $8$, we observe that ctPuLSE indeed obtains dramatically better DNSMOS OVRL scores than SuperM2M, and the enhanced target speech sounds much cleaner (please see the sound demo mentioned at the end of the Introduction section).
Although the ASR performance on the real test data is worse, it is only slightly worse and is still very strong.

\subsection{Miscellaneous Results}

Table \ref{results_speaker_reinforce} reports the results of applying speaker reinforcement (see the details in Section \ref{speaker_reinforcement}), where the SNR factor $\gamma$ is tuned to $10$ dB.
We observe that the performance gap between SuperM2M and ctPuLSE observed in Table \ref{results_comparison_with_others} is reduced, especially on the $2$- and $6$-channel tasks.
Our best performing system in row $6$b obtains ASR results competitive to using close-talk mixtures for ASR decoding, suggesting the effectiveness of ctPuLSE and the overall robust ASR system.

\ZQHL{
\section{Limitations}\label{limitation}

We emphasize that the performance of ctPuLSE depends on the quality of CTSE.
Our study builds upon the assumption that CTSE can produce reasonably-good enhancement to close-talk mixtures so that reasonably-good pseudo-labels can be derived for far-field mixtures.
This assumption, in many cases, can be satisfied as close-talk mixtures usually have a very high input SNR already.
In cases when the input SNR is low, one could leverage directional microphones to record close-talk speech during data collection. Another strategy is to leverage techniques such as cross-talk reduction \cite{Wang2024IJCAI} to train a DNN to estimate the close-talk speech based on close-talk mixtures.
}

\section{Conclusion}\label{conclusion}

We have investigated close-talk speech enhancement, and have proposed a novel approach, ctPuLSE, for far-field speech enhancement, where estimated close-talk speech produced by close-talk speech enhancement is utilized as pseudo-labels for training supervised enhancement models directly on real far-field mixtures to realize better generalizability to real data.
Evaluation results on the challenging CHiME-4 dataset show the effectiveness and potential of the proposed algorithms.

Although simple and straightforward, the proposed approach of exploiting close-talk mixtures for far-field speech enhancement, we think, could encourage a new stream of research towards realizing neural speech enhancement models that can generalize better to real data, as it suggests a promising way that can derive, for real mixtures, pseudo-labels which can enable the training of speech enhancement models directly on real mixtures.
Thanks to the innate high input SNR of close-talk mixtures, the derived pseudo-labels are often reliable and high-quality.
This paper, based on the challenging CHiME-4 dataset, has shown that the derived pseudo-labels can be utilized to build far-field speech enhancement models with better generalizability to real data.
Looking forward, we expect the derived pseudo-labels to be also useful in many other applications beyond far-field speech enhancement.

\section{Author Declarations}\label{author_declaration}

\subsection{Conflict of Interest}

The author has no conflicts to disclose. 

\section{Data Availability}

The data supporting the findings of this study is available at \url{https://catalog.ldc.upenn.edu/LDC2017S24}.

\bibliography{references.bib}

\end{document}